\documentclass[nohyper,12pt,letterpaper]{JHEP3}
\usepackage{epsfig}
\usepackage[latin1]{inputenc}
\usepackage{bbm,amsfonts}
\usepackage{graphicx}
\usepackage{amssymb,amsmath}

\author{Marco S. Bianchi,
  Matias Leoni
  and Silvia Penati \\\\
  Dipartimento di Fisica, Universit\`a di Milano--Bicocca and
  INFN, Sezione di Milano--Bicocca, Piazza della Scienza 3, I-20126 Milano, Italy
  \qquad\\\\
  E-mail: \email{marco.bianchi@mib.infn.it, matias.leoni@mib.infn.it,
    silvia.penati@mib.infn.it
    }}

\abstract{We derive an exact algebraic identity between the two--loop four--point amplitude in ABJM theory and the corresponding one--loop amplitude in ${\cal N}=4$ SYM theory. This identity generalizes previous partial results to an exact relation valid at all orders in the IR regulator. Moreover, it allows to conjecture an exact iterative expression for the complete three dimensional amplitude in terms of the BDS ansatz for the four dimensional one, indicating that the strict relation between the two amplitudes experimented at two loops might propagate to all orders. In particular, an almost complete expression for the ABJM amplitude at four loops is derived.  }

\preprint{December 2011}

\title{An all order identity between ABJM and  $\mathcal{N}=4$ SYM four--point amplitudes}


\csname @addtoreset\endcsname{equation}{section}

\def\bseq{\begin{subequation}}  
\def\eseq{\end{subequation}}
\def\bsea{\begin{subeqnarray}}  
\def\esea{\end{subeqnarray}}

\newcommand{\beq}{\begin{equation}}
\newcommand{\bea}{\begin{eqnarray}}
\newcommand{\eea}{\end{eqnarray}}
\newcommand{\eeq}{\end{equation}}

\newcommand {\non}{\nonumber}

\renewcommand{\a}{\alpha}

\newcommand{\g}{\gamma}
\newcommand{\G}{\Gamma}

\newcommand{\e}{\epsilon}
\newcommand{\z}{\zeta}

\renewcommand{\l}{\lambda}

\newcommand{\p}{\pi}

\newcommand{\Db}{\overline{D}}

\def\Mb{\kern 2pt\mathchoice
        {
         \vbox{\hrule width10pt height 0.4pt depth 0pt
         \kern 1.2pt\hbox{\kern -2pt$\displaystyle M$}}}
        {
         \vbox{\hrule width10pt height 0.4pt depth 0pt
         \kern 1.2pt\hbox{\kern -2pt$\textstyle M$}}}
        {
\vbox{\hrule width6pt height 0.4pt depth 0pt
         \kern 1.0pt\hbox{\kern -2pt$\scriptstyle M$}}}
        {
         \vbox{\hrule width5pt height 0.4pt depth 0pt
         \kern 0.8pt\hbox{\kern -2pt$\scriptscriptstyle M$}}}}

\def\Sb{\kern 2pt\mathchoice
        {
         \vbox{\hrule width6pt height 0.4pt depth 0pt
         \kern 1.2pt\hbox{\kern -2pt$\displaystyle S$}}}
        {
         \vbox{\hrule width6pt height 0.4pt depth 0pt
         \kern 1.2pt\hbox{\kern -2pt$\textstyle S$}}}
        {
         \vbox{\hrule width3.5pt height 0.4pt depth 0pt
         \kern 1.0pt\hbox{\kern -2pt$\scriptstyle S$}}}
        {
         \vbox{\hrule width3pt height 0.4pt depth 0pt
         \kern 0.8pt\hbox{\kern -2pt$\scriptscriptstyle S$}}}}

\def\Rb{\kern 2pt\mathchoice
        {
         \vbox{\hrule width5.5pt height 0.4pt depth 0pt
         \kern 1.2pt\hbox{\kern -2.5pt$\displaystyle R$}}}
        {
         \vbox{\hrule width5.5pt height 0.4pt depth 0pt
         \kern 1.2pt\hbox{\kern -2.5pt$\textstyle R$}}}
        {
         \vbox{\hrule width3.5pt height 0.4pt depth 0pt
         \kern 1.0pt\hbox{\kern -2.2pt$\scriptstyle R$}}}
        {
         \vbox{\hrule width3pt height 0.4pt depth 0pt
         \kern 0.8pt\hbox{\kern -2.2pt$\scriptscriptstyle R$}}}}

  \def\pp{{\mathchoice
      {
          \kern 1pt%
          \raise 1pt
          \vbox{\hrule width5pt height0.4pt depth0pt
            \kern -2pt
            \hbox{\kern 2.3pt
              \vrule width0.4pt height6pt depth0pt
              }
            \kern -2pt
            \hrule width5pt height0.4pt depth0pt}%
            \kern 1pt
       }
        {
          \kern 1pt%
          \raise 1pt
          \vbox{\hrule width4.3pt height0.4pt depth0pt
            \kern -1.8pt
            \hbox{\kern 1.95pt
              \vrule width0.4pt height5.4pt depth0pt
              }
            \kern -1.8pt
            \hrule width4.3pt height0.4pt depth0pt}%
            \kern 1pt
        }
        {
          \kern 0.5pt%
          \raise 1pt
          \vbox{\hrule width4.0pt height0.3pt depth0pt
            \kern -1.9pt  
            \hbox{\kern 1.85pt
              \vrule width0.3pt height5.7pt depth0pt
              }
            \kern -1.9pt
            \hrule width4.0pt height0.3pt depth0pt}%
            \kern 0.5pt
        }
        {
          \kern 0.5pt%
          \raise 1pt
          \vbox{\hrule width3.6pt height0.3pt depth0pt
            \kern -1.5pt
            \hbox{\kern 1.65pt
              \vrule width0.3pt height4.5pt depth0pt
              }
            \kern -1.5pt
            \hrule width3.6pt height0.3pt depth0pt}%
            \kern 0.5pt
        }
    }}

  \def\mm{{\mathchoice
               {
                 \kern 1pt
           \raise 1pt    \vbox{\hrule width5pt height0.4pt depth0pt
                  \kern 2pt
                  \hrule width5pt height0.4pt depth0pt}
                 \kern 1pt}
               {
                \kern 1pt
           \raise 1pt \vbox{\hrule width4.3pt height0.4pt depth0pt
                  \kern 1.8pt
                  \hrule width4.3pt height0.4pt depth0pt}
                 \kern 1pt}
               {
                \kern 0.5pt
           \raise 1pt
                \vbox{\hrule width4.0pt height0.3pt depth0pt
                  \kern 1.9pt
                  \hrule width4.0pt height0.3pt depth0pt}
                \kern 1pt}
               {
               \kern 0.5pt
         \raise 1pt  \vbox{\hrule width3.6pt height0.3pt depth0pt
                  \kern 1.5pt
                  \hrule width3.6pt height0.3pt depth0pt}
               \kern 0.5pt}
               }}

\def\pd{{\kern0.5pt
           + \kern-5.05pt \raise5.8pt\hbox{$\textstyle.$}\kern
0.5pt}}

\def\pmd{{\kern0.5pt
          \pm \kern-5.05pt
\raise6.3pt\hbox{$\textstyle.$}\kern1.5pt}}

\def\md{{\mathchoice
   {
      {{\kern 1pt - \kern-6.2pt \raise5pt\hbox{$\textstyle.$}\kern
1pt}}}
    {
      {{\kern 1pt - \kern-6.2pt \raise5pt\hbox{$\textstyle.$}\kern
1pt}}}
    {
      {\kern0.5pt - \kern-5.05pt
\raise3.4pt\hbox{$\textstyle.$}\kern0.5pt}}
    {
      {\kern0.5pt - \kern-5.05pt
\raise3.4pt\hbox{$\textstyle.$}\kern0.5pt}}}}

\def\beq{\begin{equation}}
\def\eeq{\end{equation}}
\def\bea{\begin{eqnarray}}
\def\eea{\end{eqnarray}}
\def\Tr{\textstyle{Tr}}

\def\a{\alpha}

\def\g{\gamma}

\def\e{\epsilon}
\def\z{\zeta}

\def\l{\lambda}
\def\G{\Gamma}

\renewcommand{\H}[2]{{H_{#1} #2}}

\begin{document}

\section{Introduction}

Recently, the four--point amplitude in ABJM theory \cite{ABJM} has been computed at two loops \cite{CH,BLMPS1,BLMPS2} that is the first non--trivial order where perturbative corrections appear.
This amplitude shares many remarkable properties with its analogue in ${\cal N}=4$ SYM, namely it is dual conformally invariant \cite{Drummond:2006rz,arXiv:0707.0243}, it exhibits WL/amplitude duality \cite{Brandhuber:2007yx,Drummond:2007cf} and can be consistently thought of as the first term in an exponential resummation of the perturbative series analogous to the BDS ansatz in four dimensions \cite{ABDK, BDS}. 

Actually, the amplitude itself divided by its tree level expression strikingly looks very similar to the four--gluon amplitude in ${\cal N}=4$ SYM at one--loop divided by its tree level counterpart. Precisely,  the expression of the former ratio, evaluated in dimensional regularization $d=3-2\e$, exactly matches the latter in $d=4-4\e$, up to a constant and after identifying the renormalization scales.
 
In ${\cal N}=4$ SYM theory, the IR divergent part of the amplitude is completely fixed by an evolution equation which constrains its dependence on the renormalization mass scale to be proportional to the cusp anomalous dimension $f_{\text{\tiny SYM}}(\l)$ \cite{IR}. The finite part is equally fixed by off--shell dual conformal invariance
\cite{arXiv:0707.0243,Drummond:2007cf} and the result is an iterative expression \cite{BDS}
\beq
\label{factorization}
\log\frac{\mathcal{A}_{4d}}{\mathcal{A}_{4d}^{\mathrm{tree}}} = [{\rm IR~ div}] + \frac{f_{\text {\tiny SYM}}(\l)}{4} \, 
\log^2 \left(\frac{s}{t} \right) + {\rm const.} + \mathcal{O}(\e)
\eeq
In particular, when restricted at first order in the $\l$--coupling, this identity gives the one--loop ratio on the LHS in terms of the first order expansion of the cusp anomalous dimension, as required by the Ward identities at this order.  

Given the matching between the two--loop ABJM amplitude and the one--loop ${\cal N}=4$ SYM amplitude, the Ward identities satisfied by the three dimensional amplitude are the same as those in four dimensions. Therefore, for the two--loop ratio of the three dimensional theory we expect an expansion similar to (\ref{factorization}). In fact, the explicit result found in \cite{CH,BLMPS1,BLMPS2} can be factorized as in (\ref{factorization}) and, quite remarkably, the coefficient in front of the finite remainder agrees with the ABJM cusp anomalous dimension determined through integrability arguments \cite{GV}
\beq
\label{acca}
f_{\text{\tiny CS}}(\l)=\frac12 f_{\text{\tiny SYM}}(\l)\Big|_{\frac{\sqrt{\lambda}}{4\pi } \to  h(\l)}
\eeq
being $h(\l)$ the ABJM interpolating function \cite{GV,hoflambda}.
 
Therefore, at the order we are working we can write
\beq
 \frac{\mathcal{A}^{(2)}_{3d}}{\mathcal{A}_{3d}^{\mathrm{tree}}} \Bigg|_{f_{\text{\tiny CS}}}  =
 \frac{\mathcal{A}^{(1)}_{4d}}{\mathcal{A}_{4d}^{\mathrm{tree}}} \Bigg|_{f_{\text{\tiny SYM}}} 
+ {\rm const.} + \mathcal{O}(\e) 
\eeq
where the mapping (\ref{acca}) between the two cusp anomalous dimensions is meant. 
 
Assuming that this identity can be uplifted to all orders,  we are tempted to conjecture that both amplitudes exhibit an iterative structure related by
\bea\label{logampiezze} 
\log\frac{\mathcal{A}_{3d}}{\mathcal{A}_{3d}^{\mathrm{tree}}}\Bigg|_{f_{CS}} =
\log\frac{\mathcal{A}_{4d}}{\mathcal{A}_{4d}^{\mathrm{tree}}}\Bigg|_{f_{\text{\tiny SYM}}} +\mbox{const.}+\mathcal{O}(\e)
\eea
Since the identification is up to $\mathcal{O}(\e)$ terms, the perturbative series for the two amplitudes  will not coincide.  In fact, when exponentiating eq.  (\ref{logampiezze}) to obtain the complete amplitudes the $\mathcal{O}(\e)$ terms on the RHS will mix with the $\e$--poles, spoiling the identification of the amplitudes order by order. The exact mapping between the two amplitudes can be reconstructed only once we know all the $\mathcal{O}(\e)$ terms. It is then important to check the validity of eq. (\ref{logampiezze})  to all orders in $\e$.  

In $\mathcal{N}=4$ SYM it happens that in the four--point amplitude the divergent and the non--constant finite parts in (\ref{factorization}) are completely captured by the one loop contribution in a very precise way, encoded in the well--known BDS ansatz \cite{BDS}.  Assuming that also in ABJM theory the first non--vanishing contribution to the four--point amplitude dictates all the non-trivial dependence on the kinematic invariants through its $\e$ expansion,  it is sufficient to find an all--order--in--$\e$ relation of the type (\ref{logampiezze}) only for the lowest non--trivial order in $\l$ corrections to both amplitudes. 

In this paper we carry out such a program by proving analytically an exact identity between the two--loop ABJM amplitude and the one--loop ${\cal N}=4$ SYM amplitude to all orders in $\e$. This is achieved by first 
acting on the four dimensional amplitude with a differential operator. The result can be manipulated in two different ways, so deriving two differential equations. In the first one, the RHS features the $3d$ amplitude. In the second one, a six dimensional integral appears which can be re--expressed in terms of the original $4d$ amplitude, by means of an integration by parts identity and Passarino-Veltman reduction.
Equating the RHS of these equations yields a close, all--order--in--$\e$ relation between the ABJM and ${\cal N}=4$ SYM lowest order amplitudes. Up to order $\e^2$ we have managed to check explicitly this identity by evaluating the three dimensional momentum integrals up to $\mathcal{O}(\e^2)$ and comparing the result with the known expansion of the $4d$ integrals at that order \cite{BDS}. 

The powerful identity we have found allows to rewrite the BDS--like ansatz for the ABJM four--point amplitude in terms of the four dimensional one. It follows that, if the conjectured ansatz is correct, the similarity between the two amplitudes uncovered at first order will propagate all over the perturbative expansion.  

Assuming the exponentiation ansatz to hold in three dimensions, we can speculate on the form of the four--loop amplitude. The essential ingredients determining the non--trivial parts of the four--loop correction are the cusp anomalous dimension which is known up to four loops \cite{MOS1}, and the expansion of the two--loop amplitude up to order $\e^2$. In particular, the cusp anomalous dimension along with the two--loop amplitude at order zero in $\e$, allows to fix the finite remainder of the amplitude, as shown in \cite{BLMPS1,BLMPS2}. In this paper, exploiting the knowledge of the subleading terms in $\e$, we make an almost complete prediction for the amplitude at four loops, up to constants and scheme--dependent coefficients appearing in front of simple poles.
 
The plan of the paper is as follows. In Section 2 we review the result for the ABJM four--point amplitude at two loops and discuss its similarity with the four dimensional one--loop amplitude. In Section 3 we first work out the expression of the two--loop $3d$ amplitude up to order $\e^2$, by explicitly solving the corresponding momentum integrals. This allows to realize that the matching with its four--dimensional cousin persists up to order $\e^2$. Motivated by the observation that there must be at least a technical explanation for this surprising similarity, we then give the explicit derivation of an all--order--in--$\e$ identity between the two quantities (see eq. (\ref{resultpp})). 
In Section 4, assuming the validity of a BDS--like ansatz for the $3d$ amplitude, we propose an all--loop relation between the two amplitudes. Finally, in Section 5, assuming the exponentiation ansatz to be valid,  we work out an almost complete prediction for the ABJM four--point amplitude at four loops. A final discussion and five Appendices with all technical details follow.

\section{ABJM four--point amplitude: A review}\label{sec:BDS}

In the ABJM theory, the four--point amplitude for two scalars and two fermions has been calculated at two loops in \cite{CH,BLMPS1,BLMPS2} . Its explicit expression divided by the tree level counterpart reads  
\begin{equation}\label{amplitude}
\l^2 \, \mathcal{M}^{(2)}_{3d} \equiv \frac{\mathcal{A}_4^{2-loops}}{\mathcal{A}^{tree}_4} = \lambda^2\, \left[-\frac{( s/\mu'^2)^{-2\epsilon}}{(2\, \epsilon)^2}-\frac{(t/\mu'^2)^{-2\epsilon}}{(2\, \epsilon)^2}+\frac12\,\log^2 \left(\frac{s}{t}\right)+K_1+
\mathcal{O}(\epsilon)\right]
\end{equation}
where $\lambda=N/K$ is the ABJM 't Hooft coupling (see Appendix A for conventions), $\mu'$ is the IR scale of dimensional regularization, conveniently redefined such as to absorb the $\e^{-1}$ pole
\bea\label{muprime}
\mu'^2=  8  \pi  e^{-\g_{\text{\tiny E}}}\,\mu^2
\eea
and $K_1=4\zeta_2+3\log^2 2$ is a numerical constant.

Working in ${\cal N}=2$ superspace and using an ordinary diagrammatic approach, the result (\ref{amplitude}) arises by summing contributions from six super--Feynman diagrams, after performing D--algebra reduction and computing the corresponding momentum integrals in dimensional regularization, $d= 3 -2\e$, up to order ${\cal O}(\e)$. 
We list the results for the relevant integrals, referring to \cite{BLMPS2} for a detailed explanation of their origin.  

Keeping the notation close to the one used in Refs. \cite{BLMPS1, BLMPS2},  the two--loop amplitude can be written as
\bea
\mathcal{M}^{(2)}_{3d}  = (4\pi)^2\, \left[ I^{(a)}(s) + I^{(b)}(s) + 6 I^{(d)}(s) - 2 I^{(f)}(s,t) + (s \leftrightarrow t)\right]
\eea
where
\begin{itemize}
\item Integral $(a)$
\bea
\label{Ia}
I^{(a)}(s) = -\frac{\Gamma^2(1/2+ \e) \Gamma^4(1/2 -\e)}{ (4\pi)^d \, \Gamma^2(1-2\e)} \left(\frac{\mu^2}{s}\right)^{2\epsilon} 
\eea
\item Integral $(b)$
\bea
\label{Ib}
I^{(b)}(s) = 2 \, \frac{\Gamma(1/2+\epsilon) \Gamma^2(1/2 - \epsilon) \Gamma(1+2\epsilon)\Gamma^2(-2\epsilon)}{(4\pi)^d \, \Gamma(1-2\epsilon)\Gamma(1/2-3\epsilon)}
\left(\frac{\mu^2}{s}\right)^{2\epsilon}
\eea
\item Integral $(d)$
\bea
\label{Id}
I^{(d)}(s) &=& -\frac{\Gamma^3(1/2-\epsilon)\Gamma(1+2\epsilon)\Gamma^2(-2\epsilon)}
{(4\pi)^{d} \, \Gamma^2(1-2\epsilon)\Gamma(1/2-3\epsilon)}\left(\frac{\mu^2}{s}\right)^{2\epsilon}
\eea
\item Integral $(f)$
\bea
\label{If}
&& I^{(f)}(s,t) = \frac{(1+s/t)\Gamma^3(1/2-\epsilon)}{(4\pi)^d \, \Gamma^2(1-2\epsilon)\Gamma(1/2-3\epsilon)(t/\mu^2)^{2\epsilon}} \\&&
~~~~ \times\int\limits^{+i\infty}_{-i\infty} \frac{d{\bf v}}{2\pi i}\Gamma(-{\bf v}) \Gamma(-2\epsilon -{\bf v})
\Gamma^{*}(-1-2\epsilon-{\bf v})\Gamma^2(1+{\bf v})\Gamma(2+2\epsilon+{\bf v})\left(\frac{s}{t}\right)^{{\bf v}} \non
\eea
\end{itemize}
The last expression is the result of using the Mellin--Barnes (MB) representation of the Feynman parametrized integral. The star stems for shifting the integration contour on the right of the first pole of $\Gamma(-1-2\epsilon-{\bf v})$, ensuring a well--defined expression. This integral can be evaluated in the $\e \to 0$ limit \cite{BLMPS1, BLMPS2}, leading to the finite result $(\frac12 \log^2{s/t} + 3\z_2)$.

The expression (\ref{amplitude}) is nicely akin to the one--loop four--point amplitude in ${\cal N}=4$ SYM theory \cite{4point1loop,BDDK}
\bea\label{amp4d}
\l_{\text{\tiny SYM}}\, {\cal M}_{4d}^{(1)}(\e) =  \l_{\text{\tiny SYM}} \left[-\frac{\left(s / \mu''^2 \right)^{-\epsilon }}{\epsilon ^2}-\frac{\left(t / \mu''^2 \right)^{-\epsilon }}{\epsilon ^2} + \frac12\, \log ^2\left(\frac{s}{t}\right)+\frac{2 \pi ^2}{3}\right] + {\cal O}(\e)
\eea
arising from the evaluation of a single box integral. Here, the 't Hooft coupling is defined as $\l_{\text{\tiny SYM}}=\frac{g^2 N}{8\p^2}$ and the regularization scale is
\bea\label{musecond}
\mu''^2 = 4\pi e^{-\g_{\text{\tiny E}}}\, \mu^2
\eea
Neglecting terms that vanish when removing the IR regulator, the two expressions (\ref{amplitude}) and (\ref{amp4d}) are indeed identical up to numerical constants, once we identify the scaling parameters and shift 
$\e \to 2\e$ in ${\cal M}_{4d}^{1-loop}(\e)$ to take into account the different loop order. At this stage we can then write
\beq
\label{similarity}
\mathcal{M}^{(2)}_{3d}(\e,\mu') = {\cal M}_{4d}^{(1)}(2\e,\mu'') + {\rm const.} + \mathcal{O}(\e)
\eeq
A stringent question which arises is whether the identification between the two results holds at any order in $\e$. 
The main motivation for investigating this problem comes from the observation that in ${\cal N}=4$ SYM all--order terms in the IR regulator are crucial for determining the correct exponential resummation of scattering amplitudes \cite{BDS}. Therefore, an answer to this question may shed some light on the structure of the exponentiation of scattering amplitudes in the ABJM model, as we now explain. 
 
\vskip 15pt 

Compelling evidence suggests that the four dimensional result (\ref{amp4d}) is the first order expansion of an exponential resummation of the perturbative series \cite{BDS}
\bea\label{BDS}
{\cal M}_{4d}
= \exp\left[\sum_{l=1}^\infty \l_{\text{\tiny SYM}}^l
          \left(f^{(l)}(\e) {\cal M}_{4d}^{(1)}(l \e) + C^{(l)}(\e)  \right) \right]
\eea
where ${\cal M}_{4d}^{(1)}$ is the four--point one--loop amplitude to all orders in $\e$ divided by the corresponding tree level expression, while $C^{(l)}(\e)$ contain constants plus $\mathcal{O}(\e)$ terms. 
The expansion is in powers of the dimensionless effective coupling, while the mass scale (\ref{musecond})  is hidden inside the one--loop amplitude.
The functions $f^{(l)}(\e)$ have an expansion in $\e$
\beq
f^{(l)}(\e) = f^{(l)}_0 + f^{(l)}_1 \e + f^{(l)}_2 \e^2
\eeq
whose zero order terms $f^{(l)}_0 $ coincide with one--quarter the coefficients appearing in the perturbative expansion of the scaling function $f_{\text{\tiny SYM}}$ arising in the dispersion relations for magnons \footnote{The scaling function is found to be twice the cusp anomalous dimension $\G_{cusp}$ which controls the UV divergences of Wilson loops near the cusps.}. 
 
In the ABJM theory, perturbative contributions to scattering amplitudes can occur only at even powers of
$\l= N/K$ due to the invariance of the theory under the discrete symmetry $K \rightarrow -K$, $V \leftrightarrow \hat{V}$, $A \leftrightarrow B$. Therefore, the two--loop contribution (\ref{amplitude}) is the first non--trivial quantum correction. 

This observation accompanied by the impressive similarity (\ref{similarity}) hints that the all--order amplitude for ABJM might equally exponentiate as in the four dimensional case. In other words, the amplitude (\ref{amplitude}) could be the first order expansion of
\bea\label{3dBDS}
\mathcal{M}_{3d} =  \exp\left[\sum_{l=1}^\infty \l^{2l}
          \left(\tilde{f}^{(2l)}(\e) {\cal M}_{3d}^{(2)}(l \e) + \tilde{C}^{(2l)}(\e)  \right) \right]
\eea
where the ${\cal N}=4$ SYM  coefficients $f^{(l)}(\e)$ have been replaced by their ABJM counterparts, 
$\tilde{f}^{(l)} = \tilde{f}^{(l)}_0 + \tilde{f}^{(l)}_1 \e + \tilde{f}^{(l)}_2 \e^2$ .  

A first suggestive support to this ansatz comes from the observation that, according to our two--loop calculation, at this order the coefficient $\tilde{f}^{(2)} \equiv \tilde{f}^{(2)}_0 $  matches the ABJM scaling function obtained 
through integrability in a rather independent context
\beq
\label{scaling}
\tilde{f}(\l) = \frac12 f_{\text{\tiny SYM}}(\l_{\text{\tiny SYM}}) |_{\frac{\sqrt{\l_{\text{\tiny SYM}}}}{4\pi} \to h(\l)} 
\eeq
with $h(\l)$  being the interpolating function appearing in the dispersion relations for ABJM magnons.  
 
Further support to the ansatz (\ref{3dBDS}) should come from higher order--in--$\l$ results for which 
the knowledge of the ${\cal M}_{3d}^{(2)}$ amplitude at subleading orders in $\e$ becomes mandatory.  
Moreover, according to eq. (\ref{similarity}), at two loops we can trade ${\cal M}_{3d}^{(2)}(\e)$ in eq. (\ref{3dBDS}) with ${\cal M}_{4d}^{(1)}(2\e)$.  If the identity (\ref{similarity}) were to persist at higher orders in $\e$ we could express the BDS--like ansatz for three dimensional amplitudes in terms of the four dimensional one.  Therefore, it is important to evaluate the amplitude at finite $\e$ and investigate whether the identification (\ref{similarity}) holds at any order.

\section{An identity between ABJM and ${\cal N}=4$ SYM four--point amplitudes}
 
In this Section we discuss the evaluation of subleading--in--$\e$ contributions to the four--point amplitude in ABJM theory.  

By direct inspection of $O(\e^2)$ terms we find a refinement of eq. (\ref{similarity}) which holds at that order. Thereafter, we prove a general identity between the ABJM and the ${\cal N}=4$ SYM amplitudes at all orders in the regularization parameter.  

\subsection{$O(\e^2)$ identity for the two--loop amplitudes}

When evaluating the subleading--in--$\e$ terms of the two--loop ABJM amplitude, it is more convenient to work with the IR scale $\mu''$ in eq. (\ref{musecond}) rather than $\mu'$ in eq. (\ref{muprime}). In fact, this allows to avoid the appearance of non-trivial functions of the kinematic invariants multiplying powers of $\log 2$ which would make the comparison with the four--dimensional amplitude more obscure. The price that we pay is the emergence of a simple pole divergence which modifies the relation (\ref{similarity}) as 
\beq
\label{similarity2}
\mathcal{M}^{(2)}_{3d}(\e,\mu'') = {\cal M}_{4d}^{(1)}(2\e,\mu'') - \left( s^{-2\e} + t^{-2\e} \right) \frac{1}{2\e} \log{2}  + {\rm const.} + \mathcal{O}(\e)
\eeq
To check the consistency of this relation beyond $\mathcal{O}(\e^0)$ terms, we write 
\bea\label{epsilonexp}
{\cal M}_{3d}^{(2)}(\e,\mu'') \equiv \left( s^{-2\e} + t^{-2\e} \right) \sum_{j=-\infty}^2 \frac{A_j}{\e^j}
\eea
and determine the coefficients $A_{-2}, \cdots , A_2$ by explicitly computing the integrals (\ref{Ia}--\ref{If}) up to $\e^2$. The result is reported in eq. (\ref{myresult}).

There, we also give the one--loop amplitude of ${\cal N}=4$ SYM, eqs. (\ref{shiftedpar}, \ref{modifiedBDS}), 
where we have chosen to write the result in the same form as eq. (\ref{epsilonexp}) and a doubling of the customary regularization parameter $\e$ has been performed ($d = 4 - 4\e$).

Quite remarkably, the expressions for the $A_{-2}, \cdots, A_2$ coefficients of our amplitude are almost carbon copies of the $c_{-2}, \cdots ,c_{2}$ coefficients of the one--loop amplitude in ${\cal N}=4$ SYM.
A closer look reveals that most of the differences are due to additive numerical constants which depend on the subtraction scheme that we choose.    

The only non--trivial difference between the two sets of coefficients is the appearance of a $\log^2 (s/t)$ term at order $\e^2$. 
However, it is easy to see that at least at this order we can absorb it in a scheme redefinition
\beq
\label{muA}
(\mu_A^2)^{2\e}  = [1 -5 \zeta_2 \, \e^2  + O(\e^3)] \, (\mu''^2)^{2\e} 
\eeq
thus obtaining the following empirical relation
\bea
{\cal M}_{3d}^{(2)}\left(\e,\mu_A\right) = {\cal M}^{(1)}_{4d}\left(2\e,\mu''\right) + D(\e) + {\cal O}(\e^3)
\eea
where $D(\e)$ is given in (\ref{Deps}). This result shows that the connection between the $3d$ and the $4d$ amplitudes persists at order $\e^2$. 

In the next Section we prove that this connection is not accidental but can be extended to all orders 
as an exact  identity that we derive analytically.

\subsection{All order identity for the two--loop amplitudes: An analytical derivation}\label{analytic}

First of all, we observe that writing the amplitude as in (\ref{epsilonexp}) and taking into account the particular dependence of the integrals (\ref{Ia}--\ref{If}) on the Mandelstam variables, the only non--trivial contribution to the coefficients $A_j$  comes from the integral $I^{(f)}$. In fact, all the other integrals produce  just constant factors.

Therefore, for the time being we concentrate on $I^{(f)}$.  In the following, we are going to prove an exact relationship between this integral and the four dimensional amplitude
\begin{equation}\label{resultboxed}
{\cal M}_{3d}^f(\e,\mu_A) = {\cal M}_{4d}^{(1)}(2\e,\mu'') +  \left(s^{-2\e}+t^{-2\e}\right) B(\e)
\end{equation}
where 
\beq
\label{muA2}
(\mu_A^2)^{2\e} \equiv A(\e) \, (\mu''^2)^{2\e} = 
\frac{\Gamma(1-2\e)\Gamma(1-3\e)\Gamma(1-4\e)}{\Gamma^{3}(1-\e)\Gamma(1-6\e)} \, (\mu''^2)^{2\e} 
\eeq
and $B(\e)$ is a constant given by
\bea
B(\e)=\frac{\Gamma^2(-2\e)\Gamma(1+2\e)}{\Gamma^3(1-4\e)} \, e^{2\g_{\text{\tiny E}}\e}
\eea
At second order, $\mu_A$ coincides with the expression found in (\ref{muA}). 

To prove the identity (\ref{resultboxed}) we start with the all-order-in-$\e$ expression of the integral $I^{(f)}$ as given in eq. (\ref{If}). Neglecting the mass scale (we will recover it at the very end of the derivation) and defining $x \equiv s/t$ it reads
\bea\label{MBfor3d}
{\cal M}_{3d}^f(\e) &=& \frac{(4\pi)^{2\e}\,\Gamma^3(1/2-\e)}{\pi\, \Gamma(1/2-3\e)\Gamma^2(1-2\e)} t^{-2\e}(1+x)
\non\\&&
\qquad \times \int\frac{d\bf{v}}{2\pi i}\Gamma(-v)\Gamma^2(-2\e-v)\Gamma^2(1+v)\Gamma(1+2\e+v) \, x^v
\non\\
\eea
At the same time, we consider the Mellin--Barnes representation of the $4d$ amplitude, again neglecting the mass scale. Being it given by a single box integral $I_4^{(1)}(s,t)$, we can write
\bea
\label{1loopbox}
{\cal M}_{4d}^{(1)}(2\e) &=&  - 8\, \pi^2\, s \, t \, I_4^{(1)}(s,t)  
\non \\
&=& \frac{2\e t^{-2\e}x}{(4\pi)^{-2\e}\Gamma(1-4\e)}
\int\frac{d\bf{v}}{2\pi i}\Gamma(-v)\Gamma^2(-1-2\e-v)\Gamma^2(1+v)\Gamma(2+2\e+v)\,x^v
\non \\
\eea
The contour of integration in this expression is ill--defined in the $\e\to 0$ limit, signaling the emergence of $\e$--poles. Therefore, we extract the divergent contributions by suitably deforming the contour. This leads to
\bea
&&{\cal M}_{4d}^{(1)}=
\frac{\Gamma(1+2\e)\Gamma^2(-2\e)s^{-2\e}}{2(4\pi)^{-2\e}\Gamma(-4\e)}(\log(x)+2\psi^{(0)}(-2\e)+\g_{\text{\tiny E}}-\psi^{(0)}(1+2\e))
\\&&
+\frac{2\e\, s^{-2\e}}{(4\pi)^{-2\e}\Gamma(1-4\e)}
\int\frac{d\bf{v}}{2\pi i}\Gamma(-v)\Gamma^{*\,2}(-1-2\e-v)\Gamma^2(1+v)\Gamma(2+2\e+v)\,x^{1+2\e+v}
\non 
\eea
where $\psi^{(0)}(x)$ is the digamma function defined in (\ref{polygamma}) and the Mellin--Barnes integral that survives has a well--defined contour in the $\e\to 0$ limit and contributes to the amplitude beginning at order $\e$. 

The product ${\cal M}_{4d}^{(1)}\, s^{2\e}$ depends only on the ratio $x=s/t$. Deriving with respect to $x$ we obtain
\bea\label{derivada4d}
&&\frac{d}{dx}({\cal M}_{4d}^{(1)}s^{2\e})=
\frac{\Gamma(1+2\e)\Gamma^2(-2\e)}{2(4\pi)^{-2\e}\Gamma(-4\e)}\frac{1}{x}
\\&&
+\frac{2\e\, x^{2\e}}{(4\pi)^{-2\e}\Gamma(1-4\e)}
\int\frac{d\bf{v}}{2\pi i}\Gamma(-v)\Gamma^2(-2\e-v)\Gamma^2(1+v)\Gamma(1+2\e+v)\,x^{v}\nonumber
\eea
where the same Mellin--Barnes integral as in (\ref{MBfor3d}) appears. 

Thus, comparing (\ref{derivada4d}) to (\ref{MBfor3d}) we can write
\bea\label{nexo1}
&&(1+x)\frac{d}{dx}({\cal M}_{4d}^{(1)} s^{2\e})=
-2\e \, (4\pi)^2 \, t\left[ T(s)  \, s^{2\e}  + T(t) \, t^{2\e} \right]  +\frac{2\e}{A(\e)} s^{2\e}{\cal M}_{3d}^f
\eea
where $A(\e)$ is given in eq. (\ref{muA2}) and we have defined 
\begin{equation}
\label{T}
T(s)=\frac{\Gamma(1+2\e)\Gamma^2(-2\e)}{(4\pi)^{2-2\e}\Gamma(1-4\e)}\frac{1}{s^{1+2\e}}\quad , \quad
T(t)=\frac{\Gamma(1+2\e)\Gamma^2(-2\e)}{(4\pi)^{2-2\e}\Gamma(1-4\e)}\frac{1}{t^{1+2\e}}
\end{equation}
Note that $T(s)$ and $T(t)$ are one--mass triangle integrals in four dimensions in the $s$ and $t$--channel, respectively.

Equation (\ref{nexo1}) establishes a differential relation between the $4d$ amplitude and the contribution $I^{(f)}$ to the $3d$ one. In order to obtain an algebraic relation, we derive a first order differential equation relating the four dimensional box diagram to itself.
This can be done by using an algorithm similar to the one of Ref. \cite{SV}.  

We consider the Feynman--parametrized form of the $4d$ box integral  in $d = 4-4\e$  introduced in 
(\ref{1loopbox})
\bea \label{box}
I_4^{(1)}(1,1,1,1;s,t) &=& \int\frac{d^{4-4\e}k}{(2\pi)^{4-4\e}}\frac{1}{k^2(k-p_1)^2(k-p_1-p_2)^2(k+p_4)^2}
\non \\
&=&  \frac{\Gamma(2+2\e)x^{1+2\e}}{(4\pi)^{2-2\e}} \, \frac{1}{t^{2-2\e}} \, \int\limits_{0}^{1} [d\alpha]\frac{1}{(\alpha_1\alpha_3\, x+\alpha_2\alpha_4 )^{2+2\e}}
\eea
where the measure is $[d\alpha]=d\alpha_1 d\alpha_2 d\alpha_3\,\delta(\sum_i\alpha_i-\!1)$. The labels in $I_4^{(1)}$  indicate the powers of the propagators according to the order in which they appear. 

By taking the derivative with respect to the ratio $x=s/t$ we obtain
\begin{equation}\label{derivada4d_2}
x\frac{d}{dx}(s^{1+2\e}\,t\,I_4^{(1)})=(1+2\e)s^{1+2\e}\,t\,I_4^{(1)}-
\frac{\Gamma(3+2\e)x^{2+2\e}}{(4\pi)^{2-2\e}}\int\limits_{0}^{1} [d\alpha]\frac{\alpha_1\,\alpha_3}{(\alpha_1\alpha_3\, x+\alpha_2\alpha_4 )^{3+2\e}}
\end{equation}
The second piece of this equation is proportional to a six--dimensional box integral with two indices raised by one unit. Precisely,
\begin{equation}\label{box_6}
\frac{\Gamma(3+2\e)}{(4\pi)^{3-2\e}}\int\limits_{0}^{1} [d\alpha]\frac{\alpha_1\,\alpha_3}{(\alpha_1\alpha_3\, s+\alpha_2\alpha_4 t )^{3+2\e}} = I_6^{(1)}(2,1,2,1;s,t) \equiv (\mathbf{1}^{+}\mathbf{3}^{+})I_6^{(1)}(s,t)
\end{equation}
where $\mathbf{n}^{\pm}$ are the operators which raise and lower the power of the $n$--th propagator by one unit. 

Comparing (\ref{box_6}) with (\ref{derivada4d_2}) we obtain
\begin{equation}\label{eq_con_box6}
x\frac{d}{dx}(s^{1+2\e}\,t\,I_4^{(1)}(s,t)) =(1+2\e)s^{1+2\e}\,t\,I_4^{(1)}(s,t)-4\pi\,s^{2+2\e}t\,(\mathbf{1}^{+}\mathbf{3}^{+})I_6^{(1)}(s,t)
\end{equation}

We further manipulate $(\mathbf{1}^{+}\mathbf{3}^{+})I_6^{(1)}(s,t)$ with the scope of re--expressing it in terms of four dimensional 
integrals. Applying Integration--by--parts relations arising from the identity
\begin{equation}
0=\int\frac{d^{6-4\e}k}{(4\pi)^{6-4\e}}\frac{d}{d k^{\,\mu}}
\left(\frac{(k-p_1-p_2)^{\mu}}
{k^2\,(k-p_1)^2\,[(k-p_1-p_2)^2]^2\,(k+p_4)^2}
\right)
\end{equation}
we are led to
\begin{equation}
\label{result1}
4\pi \, s\,(\mathbf{1}^{+}\mathbf{3}^{+})I_6^{(1)}(s,t)=4\pi  (\mathbf{1}^{+}+\mathbf{2}^{+}+\mathbf{3}^{+}+\mathbf{4}^{+})I_6^{(1)}(s,t)
+4\pi \, 4\e\,\mathbf{3}^{+}I_6^{(1)}(s,t)
\end{equation}
Using the Feynman--parametrized form of the first four terms, it is not difficult to ascertain that by the condition $\sum_i\alpha_i=1$ imposed by the delta--function we obtain
\begin{equation}
\label{result2}
4\pi(\mathbf{1}^{+}+\mathbf{2}^{+}+\mathbf{3}^{+}+\mathbf{4}^{+})I_6^{(1)}(s,t)=I_4^{(1)}(s,t)
\end{equation}
For the last term, comparing its Feynman--parametrized form
\begin{equation}\label{box_6_remaining}
4\pi \, 4\e\,\mathbf{3}^{+}I_6^{(1)}(s,t)=
\frac{4\e\,\Gamma(2+2\e)}{(4\pi)^{2-2\e}}\int\limits_{0}^{1} [d\alpha]\frac{\alpha_3}{(\alpha_1\alpha_3\, s+\alpha_2\alpha_4 t )^{2+2\e}}
\end{equation}
with the Feynman--parametrized form of a four--dimensional vector--like box integral
\begin{align}\label{box_vector}
I_4^{\mu}&=
\int\frac{d^{4-4\e}k}{(2\pi)^{4-4\e}}\frac{k^{\mu}}{k^2(k-p_1)^2(k-p_1-p_2)^2(k+p_4)^2}\nonumber
\\
&=\frac{\Gamma(2+2\e)}{(4\pi)^{2-2\e}}\int\limits_{0}^{1} [d\alpha]
\frac{\alpha_2 p_1^{\mu}+\alpha_3 (p_1+p_2)^{\mu}-\alpha_4 p_4^{\mu}}{(\alpha_1\alpha_3\, s+\alpha_2\alpha_4 t )^{2+2\e}}
\end{align}
we find that it coincides with the $(p_1+p_2)$--direction of the vector--like box integral in four dimensions. This component can be easily evaluated by employing Passarino--Veltman reduction and we obtain 
\begin{equation}
\label{result3}
4\pi\, 4\e\, \mathbf{3}^{+}I_6^{(1)}(s,t)=\frac{2\e}{1+x} I_4^{(1)}(s,t) +4\e\,\frac{T(t)\!-\!T(s)}{t(1+x)}
\end{equation}
where $T(s)$ and $T(t)$ have been defined in (\ref{T}).

Collecting the results (\ref{result1}, \ref{result2}, \ref{result3}) and inserting back into eq. (\ref{eq_con_box6}) we obtain
the desired differential equation for the four--dimensional box integral. Recasting it in terms of the amplitude 
$M_{4d}^{(1)}(2\e)=-8\pi^2\,s\,t\,I_4^{(1)}(s,t)$ we finally have 
\begin{equation}
(1+x)\frac{d}{dx}(M_{4d}^{(1)}(2\e)\,s^{2\e})=2\e M_{4d}^{(1)}(2\e) s^{2\e}+2\e\,(4\pi)^2\,t\,s^{2\e}[T(t)-T(s)]
\end{equation}
Comparison with the RHS of (\ref{nexo1}) produces an algebraic equation relating $M_{3d}^{f}$ to  $M_{4d}^{(1)}$ 
\begin{equation}\label{result}
M_{3d}^{f}(\e)={A(\e)}\left(M_{4d}^{(1)}(2\e)+(4\pi)^2\left(sT(s)+tT(t)\right)\right)
\end{equation}
Finally, reinserting the scale parameters and absorbing the $A(\e)(4\pi e^{-\g_{\text{\tiny E}}})^{2\e}$ factor as in (\ref{muA2}), this identity casts into the form (\ref{resultboxed}). 
 
In order to recover the whole $3d$ amplitude in terms of $M_{4d}^{(1)}$ we add to (\ref{resultboxed}) the contributions from the integrals $I^{(a)}$, $I^{(b)}$ and $I^{(d)}$ where we apply the $\mu_A$--scheme redefinition. Given that
\bea
&&\frac{(4\pi)^2}{A(\e) \, (4\pi e^{-\g_{\text{\tiny E}}})^{2\e}}\, \left( I^{(a)}(s) + I^{(b)}(s) + 6 I^{(d)}(s) + (s \leftrightarrow t)\right) = 
\non \\&\qquad&
= - \left(s^{-2\e} + t^{-2\e}\right)\, 
\frac{e^{2 \g_{\text{\tiny E}}  \epsilon }\, \Gamma (1-6 \epsilon ) \Gamma^2 \left(\frac12\,-\epsilon \right) \Gamma^3 (1-\epsilon ) }{4 \pi\, \Gamma^3 (1-2 \epsilon ) \Gamma (1-3 \epsilon )  \Gamma (1-4 \epsilon ) \Gamma \left(\frac12\,-3 \epsilon \right) }\non \\&& \hspace{10pt}
\left\{\Gamma^2 (-2 \epsilon )\Gamma (2 \epsilon +1) \left[6 \Gamma \left(\frac12\,-\epsilon \right)-2\Gamma (1-2 \epsilon ) \Gamma \left(\frac12\, + \epsilon \right)\right] \right.\non\\&&\left. \hspace{30pt}  +\Gamma \left(\frac12\,-3 \epsilon \right) \Gamma^2 \left(\frac12\,-\epsilon \right) \Gamma^2 \left(\frac12\, + \epsilon \right) \right\}
\non \\
\non \\&\qquad&
\equiv \left(s^{-2\e} + t^{-2\e}\right) E(\e)
\eea
we finally obtain \vspace{1pt}
\bea
\label{resultpp}
&& \non \\
&& \boxed{{\cal M}_{3d}^{(2)}\left(\e,\mu_A \right) =  {\cal M}_{4d}^{(1)}\left(2\e,\mu''\right) +  D(\e) } \vspace{2pt}
\\
\non 
\eea

\noindent
where we have defined $D(\e) \equiv  (s^{-2\e} + t^{-2\e}) (B(\e) + E(\e))$. 
Series expansions for $A(\e)$, $B(\e)$, $D(\e)$ and $E(\e)$ can be found in Appendix D. It is  easy to check that neglecting subleading terms for $\e \to 0$ we are back to the relation (\ref{similarity2}).

\section{Three dimensional BDS ansatz revisited}\label{sec:3dBDS}

The identity (\ref{resultpp}) enables us to reformulate the BDS--like conjecture for the ABJM four--point amplitude in terms of the original ${\cal N}=4$ SYM all--loop proposal.

In fact, reformulating the ansatz (\ref{3dBDS}) in the $\mu_A$ scheme, first of all we can write   
\bea
\label{3dBDS2}
\mathcal{M}_{3d} (\e,\mu_A)=  \exp\left[\sum_{l=1}^\infty \l^{2l}
          \left(f_{\text{\tiny CS}}^{(2l)}(\e) {\cal M}_{3d}^{(2)}(l \e,\mu_A) + {C}_{\text{\tiny CS}}^{(2l)}(\e)  \right) \right]
\eea
where the functions $f^{(2l)}_{\text{\tiny CS}}(\e)$ are the $\tilde{f}^{(2l)}(\e)$ counterparts in the $\mu_A$--scheme. Their leading coefficients $f_{\text{\tiny CS},0}^{(2l)}$ are still determined by one--quarter the Chern--Simons scaling function, as the change of scheme affects only $\tilde{f}_{2}^{(2l)}$. 

Modifications in the constant part of the amplitude due to the scheme change are included in the new coefficients ${C}_{\text{\tiny CS}}^{(2l)}(\e)$. 

The convenient choice of the $\mu_A$--scheme allows to use the identity (\ref{resultpp}) in the previous expression, thus leading to a suggestive ansatz for the all--loop four--point amplitude in $3d$
\bea
\label{BDSrevisited}
{\cal M}_{3d}(\e, \mu_A) &=& \exp \left[ \sum_{l=1}^\infty \l^{2l}
          \left(f_{\text{\tiny CS}}^{(2l)}(\e) \, {\cal M}_{4d}^{(1)}(2l \e,\mu'') + f_{\text{\tiny CS}}^{(2l)}(\e) \, D(l\e) + C_{\text{\tiny CS}}^{(2l)}(\e)  \right)  \right] \non\\
              &\equiv& {\cal M}_{4d}(2\e, \mu'') \Big|_{f(\e)\rightarrow f_{\text{\tiny CS}}(\e)} \exp \left[ \sum_{l=1}^\infty \l^{2l}
            {\cal H}^{(l)}(\e) \right]
\eea
where we have defined 
\beq
{\cal H}^{(l)}(\e) = f^{(2l)}_{\text{\tiny CS}}(\e)\, D(l\e) + C_{\text{\tiny CS}}^{(2l)}(\e) -  C^{(l)}(2\e) 
\eeq
and $C^{(l)}(2\e)$  are the functions appearing in the $4d$ BDS ansatz (\ref{BDS}).

It is important to note that, apart from a factor $(s^{-2l\e} + t^{-2l\e})$ hidden inside $D(l\e)$, the coefficients ${\cal H}^{(l)}(\e)$ contain only constant terms, while the hard--core of the amplitude is completely encoded in ${\cal M}_{4d}$.
 
The ansatz (\ref{BDSrevisited}) reveals a deep intertwining between the conjectured exponentiation of four--point amplitudes in ${\cal N}=4$ SYM and ABJM theories. Therefore, we are led to conjecture that the remarkable connection uncovered at lowest order \cite{CH,BLMPS1,BLMPS2} will propagate over their entire perturbative series.

\section{A conjecture for the four--loop amplitude}\label{sec:prediction}

In spite of this beautiful result, the poor knowledge of the functions $f_{\text{\tiny CS}}(\e)$ spoils the power of (\ref{BDSrevisited}) in predicting higher order corrections to the four--point function.
Nevertheless, using known results for the scaling function of the ABJM theory, we are able to formulate an almost complete prediction for this amplitude at four loops in terms of the ${\cal N}=4$ SYM  amplitude at two loops. 
 
The main ingredients for carrying out this program are the expression 
\bea
\label{4loops1}
 {\cal M}_{3d}^{(4)}\left(\e,\mu_A\right) &=& \frac12 \left[   f_{\text{\tiny CS}}^{(2)}(\e) \left( {\cal M}_{4d}^{(1)}(2\e,\mu'') + D(\e) \right) + C_{\text{\tiny CS}}^{(2)}(\e) \right]^2  
\non \\
&& \qquad +  f_{\text{\tiny CS}}^{(4)}(\e)   \left( {\cal M}_{4d}^{(1)}(4\e,\mu'') + D(2\e) 
+ C_{\text{\tiny CS}}^{(4)}(\e) \right)
\eea
obtained from the expansion of the ansatz (\ref{BDSrevisited}), and the value of the ABJM scaling function at four--loops \cite{MOS1}
\beq
f_{\text{\tiny CS}}(\lambda)= 4 \lambda^2 - 24\, \zeta_2\, \lambda^4 + {\cal O}(\lambda^6)
\eeq
which leads to
\bea
\label{4loopsf}
f^{(2)}_{\text{\tiny CS}} \equiv f^{(2)}_{\text{\tiny CS},0} = 1 \qquad f^{(4)}_{\text{\tiny CS},0} = -6\, \zeta_2
\eea
Unfortunately, while integrability suggests a prescription for deriving the ABJM scaling function from that of $\mathcal{N}=4$, no such a connection is known for the first and the second order coefficients $f_{\text{\tiny CS},1}^{(4)}$ and $f_{\text{\tiny CS},2}^{(4)}$. 

From the structure of the BDS--like ansatz it is easy to realize that as long as we are interested in the non--trivial part of the amplitude we can forget about $f_{\text{\tiny CS},2}^{(4)}$ that would contribute only to constants. 
The lack of information about $f_{\text{\tiny CS},1}^{(4)}$, instead, leaves the $1/\e$ pole undetermined \footnote{To restrain the lack of information coming from the unknown $f_{\text{\tiny CS},1}^{(4)}$, one could switch to a scheme where the choice of the regularization scale would be analogous to that in (\ref{muprime}) ($\mu_A^2\rightarrow \mu_A'^2=2^{2\e} \mu_A^2$), allowing for no $1/\e$ poles in the series expansion of the two--loop amplitude. In that case $f_{\text{\tiny CS},1}^{(4)}$ would only affect the coefficient of the $1/\e$ pole at four loops, but not the finite part.}. Therefore, the two relations above are sufficient to formulate an almost complete prediction, up to scheme--dependent subdivergent terms.

Combining the two ingredients (\ref{4loops1}, \ref{4loopsf}), we can write 
\bea
{\cal M}_{3d}^{(4)}\left(\e,\mu_A\right) &=& \frac12 \left( {\cal M}_{4d}^{(1)}(2\e,\mu'') + D(\e) \right)^2  - 6\, \zeta_2\, \left( {\cal M}_{4d}^{(1)}(4\e,\mu'') + D(2\e) \right)\non\\&& - \left(s^{-4\e} + t^{-4\e}\right)\frac{f^{(4)}_{\text{\tiny CS},1}}{16\e} + \hat{C}^{(4)}_{\text{\tiny CS}}(\e)
\eea
where $\hat{C}^{(4)}_{\text{\tiny CS}}(\e)$ includes finite contributions coming from $f^{(4)}_{\text{\tiny CS},1}$ and $f^{(4)}_{\text{\tiny CS},2}$. 

Alternatively, we can collect the pieces that reproduce the ${\cal N}=4$ SYM amplitude at two loops \cite{BDK2loops} and cast
the previous expression into the form 
\bea\label{4L3d4d}
{\cal M}_{3d}^{(4)}\left(\e,\mu_A\right) &=& {\cal M}_{4d}^{(2)}(2\e,\mu'') +  {\cal M}_{4d}^{(1)}(2\e,\mu'')  D(\e)- 5\, \zeta_2\, {\cal M}_{4d}^{(1)}(4\e,\mu'') 
\non\\&& 
- 6\, \zeta_2\, D(2\e) + \frac12 D(\e)^2  
- \left(s^{-4\e} + t^{-4\e}\right)\frac{\hat{f}^{(4)}_{\text{\tiny CS},1}}{16\e} + \hat{\hat{C}}^{(4)}_{\text{\tiny CS}}(\e) \non\\
\eea
where $\hat{f}^{(4)}_{CS,1}$ and $\hat{\hat{C}}^{(4)}_{\text{\tiny CS}}$ have been defined so as to include ${\cal O}(\e^{-1})$ and ${\cal O}(\e^0)$ terms respectively, arising when reconstructing the four dimensional four loop amplitude\footnote{More precisely $\hat{f}^{(4)}_{\text{\tiny CS},1} = f^{(4)}_{\text{\tiny CS},1} - 2\, f^{(2)}_{\text{\tiny SYM},1}\, \left(-\frac{1}{16}\right) = f^{(4)}_{\text{\tiny CS},1} - \frac{1}{8}\, \zeta_3 $, and $\hat{\hat{C}}^{(4)}_{\text{\tiny CS}} = \hat{C}^{(4)}_{\text{\tiny CS}} - 4\, f^{(2)}_{YM,2}\, \left(-\frac{1}{16}\right) = \hat{C}^{(4)}_{\text{\tiny CS}} - \frac14\, \zeta_3 $. }. 

Eq. (\ref{4L3d4d}) expresses our prediction for the four--loop four--point amplitude in ABJM in terms of the two--loop amplitude in ${\cal N}=4$ SYM theory. An explicit expression for its divergent and finite parts is given in Appendix E.  In the $C_0$ coefficient, eq. (\ref{4loops}), we can recognize terms proportional to $L^4$, 
$(\log^2 2) L^2$ and $\z_2 L^2$ which were present in the finite reminder computed in \cite{BLMPS2}.  

We stress that the most non--trivial contribution to the amplitude, featuring the highest weight harmonic polylogarithms, is enclosed in the four dimensional two--loop contribution (the first term on the RHS of this equation). Moreover, all the remaining polylogarithmic dependence on the kinematic invariants is captured by the four--dimensional lower order amplitudes. This pattern will show up at any perturbative order.  

\section{Discussion}

In this paper we have algebraically derived an exact relation between the two--loop four--point amplitude of the ABJM theory divided by its tree level expression and the corresponding one--loop ratio in the ${\cal N}=4$ SYM theory. That a relation were to be at work was already observed in \cite{CH, BLMPS1, BLMPS2} for the two amplitudes computed up to finite terms in the IR regulator. Here, we have extended this relation to an exact identity holding at any order in $\e$. 

Besides the technical reasons underlying this identity which can be inferred from our proof, it would be very interesting to understand whether it has a deeper explanation based on more robust conceptual grounds. Moreover, still in this direction, it would be important to investigate whether it is an accident of the four--point amplitude or it holds also for higher--point amplitudes. 

Assuming that, as in the ${\cal N}=4$ SYM theory, the ABJM amplitudes have an iterative structure encoded in a BDS--like exponentiation ansatz, the all--order--in--$\e$ result at lowest order in perturbation theory would be sufficient for identifying all higher order contributions.  In three dimensions this is indeed two loops. According to our identity, the two--loop $3d$ amplitude can be rewritten in terms of the lowest order contribution to the four dimensional one which enters the ordinary BDS 
equation for the whole $4d$ amplitude. Therefore, this allows to conjecture that an exact relation should hold between the 
complete four--point three dimensional amplitude and its four dimensional counterpart. 

A strong non--trivial test of this conjecture would come from a direct evaluation of the ABJM amplitude at four loops, either 
by traditional perturbative methods or by generalized unitarity cuts. The result could in fact confirm or kill the prediction that we have 
for the amplitude at this order and, at the same time, could give some indication on the explicit expression for the scheme dependent coefficient $f^{(4)}_{\text{\tiny CS},1}$ that our ansatz leaves undetermined. 

In Refs. \cite{CH, BLMPS1, BLMPS2} a duality between the two--loop four--point amplitude and a $3d$ light--like four--polygon Wilson loop \cite{Henn:2010ps, Wiegandt:2011uu} was pointed out, as long as the two quantities are evaluated up to finite terms in $\e$. It would be interesting to investigate whether this duality persists at sub--leading orders. Given our algebraic identity and the fact that the $4d$ amplitude is known up to order $\mathcal{\e}^4$ \cite{BDS}, the explicit result for the two--loop ABJM amplitude at that order in $\e$ is now available. It would then be interesting to try and push the evaluation of the Wilson loop up to the same order and compare the two results.

\vskip 25pt
\section*{Acknowledgements}
\noindent
We thank Andrea Mauri and Alberto 
Santambrogio for valuable discussions.
This work has been supported in part by INFN and by MIUR--PRIN contract 2009-KHZKRX.

\vfill
\newpage

\appendix

\section{ABJM theory: Notations and conventions}

In ${\cal N}=2$ superspace, the physical content of $U(N)_{K} \times U(N)_{-K}$ ABJM theory \cite{ABJM} is organized into two vector multiplets $(V,\hat{V})$ in the adjoint representation of the first and the second $U(N)$'s respectively, and
four chiral multiplets $A^i$ and $B_i$, $i=1,2$, with $A^i$ in
the $(N,\bar{N})$ and $B_i$ in the $(\bar{N},N)$ (anti)bifundamental
representations.

The ${\cal N}=6$ supersymmetric action reads \cite{Klebanov,BPS1}
\begin{equation}
 {\cal S} = {\cal S}_{\mathrm{CS}} + {\cal
    S}_{\mathrm{mat}}
  \label{eqn:action}
\end{equation}
with
\begin{eqnarray}
  \label{action}
  && {\cal S}_{\mathrm{CS}}
  =  \frac{K}{4\pi} \, \int d^3x\,d^4\theta \int_0^1 dt\: \Big\{  \Tr \Big[
  V \Db^\a \left( e^{-t V} D_\a e^{t V} \right) \Big]
  -   \Tr \Big[ \hat{V} \Db^\a \left( e^{-t \hat{V}} D_\a
    e^{t \hat{V}} \right) \Big]   \Big\}
  \non \\
  \non \\
  && {\cal S}_{\mathrm{mat}} = \int d^3x\,d^4\theta\: \Tr \left( \bar{A}_i
    e^V A^i e^{- \hat{V}} + \bar{B}^i e^{\hat V} B_i
    e^{-V} \right)
  \non \\
  &~& ~ + \frac{2\pi i}{K} \int d^3x\,d^2\theta\:
    \e_{ik} \e^{jl}  \, {\Tr (A^i B_j A^k B_l)} +  \frac{2\pi i}{K} \int d^3x\,d^2\bar{\theta}\:
  \e^{ik} \e_{jl} \, {\Tr (\bar{A}_i \bar{B}^j \bar{A}_k\bar{B}^l}  )
  \non\\
\end{eqnarray}
where $K$  is an integer, as required by the gauge invariance of the effective action.
In the perturbative regime we take $\l \equiv \frac{N}{K} \ll 1$.

We are interested in four--point scattering amplitudes. Without loosing generality we consider
chiral superamplitudes  of the type $(A^i B_j A^k B_l)$, as the other superamplitudes  can be obtained from these ones by $SU(4)$ R--symmetry transformations.

At any order in perturbation theory they can be inferred from the corresponding contributions to the effective superpotential \cite{BLMPS1,BLMPS2}. When going to components, they give rise to amplitudes for two scalars and two fermions. 

We write the amplitude divided by its tree level expression as an expansion in powers of the 't Hooft coupling 
\beq
{\cal M} = \sum_l \l^l\, {\cal M}^{(l)}
\eeq
where $\l$ is the dimensionless effective coupling in $d=3-2\e$, and 
${\cal M}^{(l)} \equiv {\cal A}^{(l-loops)}/{\cal A}^{tree} $ includes the mass scale of dimensional regularization $(\mu^2)^{l\e}$.

\section{Functions appearing in the result at order $\e^2$}

Here we review the definition of harmonic polylogarithms $H_{a_1\dots a_n}$ which are ubiquitous in solving Feynman integrals.
These functions are defined recursively as
\begin{equation}
H_{a_1 a_2 \ldots a_n}(x)
= \int_0^x d t \,  f_{a_1}(t) H_{a_2 \ldots a_n}(t)
\end{equation}
where
\begin{eqnarray}
f_{\pm 1}(x) &=& \frac{1}{1 \mp x} \,, \qquad f_{0}(x) = \frac{1}{x} \,, \\
H_{\pm 1}(x) &=& \mp \log(1\mp x) \,, \qquad H_{0}(x)= \log x \,,
\end{eqnarray}
and the index $n$ is referred to as the weight of the given harmonic polylogarithm.
Up to weight 4, these functions (when only 0 or +1 indices are present) may be expressed in terms of ordinary polylogarithms.

All the harmonic series arising from closing our MB integrals can be solved analytically.
Simple and nested harmonic sums are conventionally defined as \cite{Smirnov}
\bea
S_i(n) = \sum_{j=1}^n \frac{1}{j^i} \qquad S_{i,k}(n) = \sum_{j=1}^n \frac{S_k(j)}{j^i}
\eea
Here we list some nested series, which are encountered during the order $\e^2$ evaluation of integrals, producing high order harmonic polylogarithms
\bea
&& \sum_{n=1}^{\infty} S_1(n-1)^2 \frac{z^n}{n} = -2S_{1,2}(z) + \H{1}{(z)} \H{0,1}{(z)} + \frac13 \H{1}{(z)}^3   \\&&
\sum_{n=1}^{\infty} S_1(n-1) S_2(n-1) \frac{z^n}{n} = -\frac12 \H{0, 1}{(z)}^2 - \H{1}{(z)} \left( S_{1,2}(z) - \H{0,0,1}{(z)} \right) +  \non\\&& ~~~~~~~~~~~~~~~~~~~~~~~~~~~~~~~~~~~~~~~ \frac12 \H{1}{(z)}^2 \H{0,1}{(z)}^2  \\&&
\sum_{n=1}^{\infty} S_2(n-1)^2 \frac{z^n}{n} = \H{1, 0, 0, 0, 1}{\left(z\right)} + 2 \H{1, 0, 1, 0, 1}{\left(z\right)}
\eea
These are relevant for summing series of second order in polygamma functions $\psi^{(i)}(n)$
\beq\label{polygamma}
\psi^{(i)}(x) = \frac{d^{i+1}}{dx^{i+1}}\, \log \Gamma\left( x \right)  
\eeq
such as $\sum_{n=1}^{\infty} \left[\psi^{(0)}(n)\right]^2 z^n$, $\sum_{n=1}^{\infty} \psi^{(0)}(n) \psi^{(1)}(n) z^n$ and $\sum_{n=1}^{\infty} \left[\psi^{(1)}(n)\right]^2 z^n$, respectively.

\section{Explicit Comparison}

In this section we compute explicitly the three dimensional amplitude at order $\e^2$ and compare it to the four dimensional one.

In order to make closer contact with the four dimensional case we choose the IR scale to be $\mu''^2 = 4\p e^{-\g_{\text{\tiny E}}}\, \mu^2$.
This choice has the advantage that many non--trivial terms multiplied by powers of $\log 2$ drop off the series. The price to pay is to reinstate a $\e^{-1}$ pole, which could have been removed from the coefficients $A_j$ by using the scale $\mu'$ in (\ref{muprime}).
We write
\bea\label{ampgenmusecond}
{\cal M}_{3d}^{(2)}(\e,\mu'') = \left( s^{-2\e} + t^{-2\e} \right) \sum_{j=-2}^2 \frac{A_j}{\e^j}
\eea
By $\e$--expanding the integrals arising in the computation and using the tools reviewed above we determine explicitly the coefficients.
With the shorthands $x \equiv s/t$ and $L \equiv \log x$, they read
\bea\label{myresult}
A_{2} & = & -\frac14 \non\\
A_{1} & = & -\frac12 \log{2} \non\\
A_{0} & = & +\frac14 L^2 + \frac{\pi ^2}{3} + \log^2{2} \non\\
A_{-1} & = & -\frac12 \H{0,0,1}{\left(-\frac{1}{x}\right)} - \frac12 \H{0,1}{\left(-\frac{1}{x}\right)} L \non\\&&
- \frac14 \H{1}{ \left(-\frac{1}{x}\right)} L^2 - \frac14 \pi ^2 \H{1}{\left(-\frac{1}{x}\right)} + \left(x\leftrightarrow \frac{1}{x}\right) \non\\&& + \frac{185 \zeta (3)}{12}-\frac{4 \log^3{2}}{3}-\frac{7}{12} \pi ^2 \log{2} \non\\
A_{-2} & = &   
+ \H{1, 0, 0, 1}{\left(-\frac{1}{x}\right)}
  + \H{0, 0, 1, 1}{\left(-\frac{1}{x}\right)}
  + \H{0, 1, 0, 1}{\left(-\frac{1}{x}\right)}
  + \H{0, 0, 0, 1}{\left(-\frac{1}{x}\right)}
\non \\&&
+  H_{0,1,1}\left(-\frac{1}{x}\right) L + H_{1,0,1}\left(-\frac{1}{x}\right) L
\non \\&&
-\frac12\,   H_{0,1}\left(-\frac{1}{x}\right)L^2  + \frac12\, H_{1,1}\left(-\frac{1}{x}\right) L^2 + \frac12\, \pi ^2 H_{1,1}\left(-\frac{1}{x}\right) 
\non \\&&
-\frac{1}{3} H_1 \left(-\frac{1}{x}\right) L^3 - \frac12\, \pi ^2 H_1 \left(-\frac{1}{x}\right) L -   \zeta_3 \,H_1\left(-\frac{1}{x}\right)  + \left(x\leftrightarrow \frac{1}{x}\right)
\non \\&&
- \frac{1}{24} \pi ^2 L^2
\non \\&&
+\frac{91 \zeta (3) \log{2}}{3}+\frac{233 \pi ^4}{1440}+\frac{4 \log^4{2}}{3}-\frac{17}{6} \pi ^2 \log^2{2}
\eea

In order to make a comparison, we report the corresponding result for the one--loop four--point amplitude in ${\cal N}=4$ SYM as can be found in Ref. \cite{BDS}. Using the results of Appendix B of that reference, suitably adapted to our conventions (euclidean signature, $x \to 1/x$ and $\e \to 2\e$) and choosing a slightly different definition for the series expansion  
\bea\label{shiftedpar}
{\cal M}_{4d}^{(1)}(\e,\mu'') = \left( s^{-2\e} + t^{-2\e} \right) \sum_{j=-2}^2 \frac{c_j}{\e^j}
\eea
we find
\bea\label{modifiedBDS}
c_2 & = & -\frac14
\non \\
c_1 &=& 0
\non \\
c_0 & = &
+\frac{L^2}{4}+\frac{\pi ^2}{3}
\non \\
c_{-1} & =&
 - \frac12\H{0, 0, 1}{\left(-\frac{1}{x}\right)} - \frac12 \H{0, 1}{\left(-\frac{1}{x}\right)} L \non\\&& 
- \frac14 \H{1}{\left(-\frac{1}{x}\right)} L^2 -\frac14 \pi^2 \H{1}{\left(-\frac{1}{x}\right)} + \left(x\leftrightarrow \frac{1}{x}\right) \non\\&& 
 +  \frac{17}{3} \zeta_3 
\non \\
c_{-2} & =&
  + \H{1, 0, 0, 1}{\left(-\frac{1}{x}\right)}
  + \H{0, 0, 1, 1}{\left(-\frac{1}{x}\right)}
  + \H{0, 1, 0, 1}{\left(-\frac{1}{x}\right)}
  + \H{0, 0, 0, 1}{\left(-\frac{1}{x}\right)}
\non \\
&& \null
  + \H{0, 1, 1}{\left(-\frac{1}{x}\right)} L
  + \H{1, 0, 1}{\left(-\frac{1}{x}\right)} L
\non \\
&& \null
  - \frac12\H{0, 1}{\left(-\frac{1}{x}\right)} L^2
  + \frac12\H{1, 1}{\left(-\frac{1}{x}\right)} L^2
  + \frac12\pi^2 \H{1, 1}{\left(-\frac{1}{x}\right)}
\non \\
&& \null
  - \frac13 \H{1}{\left(-\frac{1}{x}\right)} L^3
  - \frac12\pi^2 \H{1}{\left(-\frac{1}{x}\right)} L
  - \zeta_3  \, \H{1}{\left(-\frac{1}{x}\right)}   + \left(x\leftrightarrow \frac{1}{x}\right)
\non \\
&& \null
  + \frac13 \pi^2 L^2
  + \frac{20}{3} \zeta_3 \, L + \frac{41}{360}  \pi^4
\eea
Comparing them with eq. (\ref{myresult}) we see that the two sets of coefficients differ by constants and by a non--trivial term $-\frac{5}{24} \pi ^2 L^2 = -\frac{5}{4} \zeta_2 \log^2{(s/t)}$ at order $\e^2$.
Owing to this observation we conclude that up to order $\e^2$ the following relation holds
\beq\label{finalrel}
\frac{{\cal M}_{3d}^{(2)}(\e)}{A(\e)} = {\cal M}_{4d}^{(1)}(2\e) +  D(\e) + {\cal O}(\e^3)
\eeq
where $A(\e)= 1 -5 \zeta_2 \e^2 + O(\e^3)$ and $D(\e)$ is given in (\ref{Deps}). This precisely meets the expectations from (\ref{resultpp}), which proves this formula to be valid to all orders in $\e$.

\section{Series expansions}

Here we list the $\e$--expansions of the functions used in the computations of Sections \ref{analytic}, \ref{sec:3dBDS} and \ref{sec:prediction}
\bea\label{Aeps2}
A(\e)&=&\frac{\Gamma(1-2\e)\Gamma(1-3\e)\Gamma(1-4\e)}{\Gamma^{3}(1-\e)\Gamma(1-6\e)}\non\\&=&
1-5\zeta_2\,\e^2-40\zeta_3\,\e^3-\frac{821 \zeta_4}{4}\, \epsilon ^4
+\mathcal{O}(\e^5)
\eea
\bea\label{Beps2}
B(\e)&=&\frac{\Gamma^2(-2\e)\Gamma(1+2\e)}{\Gamma^3(1-4\e)} \, e^{2\g_{\text{\tiny E}}\e}\non\\&=&
\frac{1}{4 \epsilon ^2}-\frac12\, \zeta_2-\frac{14}{3}\,  \zeta_3\, \epsilon -\frac{47}{4}\, \zeta_4\, \epsilon^2 + {\cal O}(\e^3)
\eea
\bea\label{Eeps}
E(\e) &=& 
-\frac{1}{4 \epsilon ^2} - \frac{\log 2}{2 \epsilon } + \left(\log ^2 2 - \frac34\, \zeta_2\right)\non\\ && + \left(-6\, \zeta_2\, \log 2+\frac{53}{12}\, \zeta_3 - \frac43\, \log ^3 2 \right) \e \non\\ && + \left(-12\, \zeta_2\, \log ^2 2 - \frac{207}{8}\, \zeta_4 + \frac{31}{3}\, \zeta_3\, \log 2 + \frac43\, \log ^4 2 \right) \e^2 + {\cal O}\left(\e^3\right) \non\\&&
\eea
\bea\label{Deps}
D(\e) &=& \left(s^{-2\e} + t^{-2\e}\right)\, \left\{
-\frac{1}{2 \epsilon } \log 2+ \left(\log ^2 2-\frac54\, \zeta_2\right) \right. 
\non\\&& 
\left. + \left(-6\, \zeta_2\, \log 2-\frac14\, \zeta_3 - \frac43\, \log ^3 2 \right) \e \right.
\non\\&& 
\left. + \left(-12\, \zeta_2\, \log ^2 2-\frac{301}{8}\, \zeta_4 + \frac{31}{3}\, \zeta_3\, \log 2 + \frac43\, \log ^4 2 \right) \e^2 + {\cal O}\left(\e^3\right) \right\} \non\\&&
\eea

\section{The ABJM four--point amplitude at four loops}

In this Appendix we give the explicit result for the ABJM four--point amplitude at four loops as derived from eq. (\ref{4L3d4d}). Writing
\beq
  {\cal M}_{3d}^{(4)}\left(\e,\mu_A\right) = \left( s^{-4 \e} + t^{-4 \e} \right)\, \sum_{j=0}^4 \frac{C_j}{\e^j} + {\cal O}(\e)
\eeq	
we find
\bea
\label{4loops}
C_4 &=& \frac{1}{16}
\non \\
C_3 &=& \frac14 \log 2
\non \\
C_2 &=& -\frac{3}{16}\, L^2 - \frac14 \log ^2 2
\non \\
C_1 &=&  \frac14\, \H{0, 0, 1}{(-x)} 
 - \frac14 \H{0,1}{(-x)}\, L 
\non \\
&& \null
+ \frac18 \, \H{1}{(-x)} \, L^2
+ \frac18\, \pi ^2\, \H{1}{(-x)}
+ \left( x\leftrightarrow \frac{1}{x} \right) 
 - \frac12  \, \log 2 \, L^2
\non \\
&& \null
 - \frac{65}{24}\, \zeta_3 - \frac13\, \log ^3 2 + \frac{15}{4}\, \zeta_2\, \log 2
-\frac{f^{(4)}_{\text{\tiny CS},1}}{16}
\non \\
C_0 &=& 
- \frac12\, \H{0,0,0,1}{(-x)} 
- \frac12\, \H{0, 0,1, 1}{(-x)}
- \frac12\, \H{1,0,0,1}{(-x)}
 - \frac12\, \H{0, 1,0, 1}{(-x)} 
\non \\
&& \null
+ \frac12 \, \H{0,1, 1}{(-x)} \, L
+ \frac12 \, \H{1,0,1}{(-x)} \, L
-\frac14 \, \H{1, 1}{(-x)} \, L^2
\non \\
&& \null
+ \frac14 \, \H{0,1}{(-x)} \, L^2
-\frac32\, \zeta_2\, \H{1, 1}{(-x)} 
+\frac12\, \zeta_3\, \H{1}{(-x)}
\non \\
&& \null
+ \frac{1}{12}\, \log 2\, \Big( 
 6\, \H{0, 0, 1}{(-x)}
- 6\, L\, \H{0,1}{(-x)}
\non\\&& \null ~~~~~~~~~~~~~~~~ 
+ 3 \, \H{1}{(-x)} \, L^2
+ 3\, \pi ^2\, \H{1}{(-x)} \Big)
+ \left( x\leftrightarrow \frac{1}{x} \right) 
\non \\
&& \null
+ \frac{11}{24}\, L^4 
+\frac34\, \log ^2 2 \, L^2 + \frac14\, \zeta_2\, L^2
\non \\
&& \null
+ \frac{123}{32}\, \zeta_4 
- \frac{127}{12}\, \zeta_3\, \log 2 + \frac{15}{2}\, \zeta_2\, \log ^2 2 + \frac53\, \log ^4 2
\non \\
&& \null
- \frac14\, \log 2\, f^{(4)}_{\text{\tiny CS},1} - \frac{1}{16}\, f^{(4)}_{\text{\tiny CS},2} + \frac12\, C^{(4)}_{\text{\tiny CS}}(0)
\eea

\end{document}